\documentclass[10pt,twocolumn,letterpaper]{article}
\usepackage{makecell}
\usepackage{times}
\usepackage{epsfig}
\usepackage{graphicx}
\usepackage{graphics}
\usepackage{amsmath}
\usepackage{amssymb}
\usepackage{booktabs, multirow}
\usepackage{amsmath,amssymb,amsthm}
\usepackage{algorithm}
\usepackage{algorithmic}
\usepackage{bm}
\usepackage{enumerate}
\usepackage{standalone}
\newtheorem{thm}{Theorem}
\newtheorem{lemma}{Lemma}

\theoremstyle{definition}

\newtheorem{remark}{Remark}

\newcommand*{\affaddr}[1]{#1} 
\newcommand*{\affmark}[1][*]{\textsuperscript{#1}}
\newcommand*{\email}[1]{\texttt{#1}}

\usepackage[breaklinks=true,bookmarks=false]{hyperref}

\begin{document}

\title{Synthetic Learning: Learn From Distributed Asynchronized Discriminator GAN Without Sharing Medical Image Data}

\author{%
Qi Chang\affmark[1]\thanks{equal contribution}, Hui Qu\affmark[1]\footnotemark[1], Yikai Zhang\affmark[1]\footnotemark[1], Mert Sabuncu\affmark[2], \\Chao Chen\affmark[3], Tong Zhang\affmark[4] and Dimitris Metaxas\affmark[2]\\
\affaddr{\affmark[1]Rutgers University}
\affaddr{\affmark[2]Cornell University}\\
\affaddr{\affmark[3]Stony Brook University}
\affaddr{\affmark[4]Hong Kong University of Science and Technology}\\
\email{\{qc58,hq43,yz422,dnm\}@cs.rutgers.edu} , \email{msabuncu@cornell.edu},\\ \email{chao.chen.cchen@gmail.com}, \email{tongzhang@tongzhang-ml.org}%
}


\maketitle

\begin{abstract}
In this paper, we propose a data privacy-preserving and communication efficient distributed GAN learning framework named Distributed Asynchronized Discriminator GAN (AsynDGAN). 
Our proposed framework aims to train a central generator learns from distributed discriminator, and use the generated synthetic image solely to train the segmentation model.
We validate the proposed framework on the application of \emph{health entities learning problem} which is known to be privacy sensitive. 
Our experiments show that our approach: 1) could learn the real image's distribution from multiple datasets without sharing the patient's raw data. 2) is more efficient and requires lower bandwidth than other distributed deep learning methods. 3) achieves higher performance compared to the model trained by one real dataset, and almost the same performance compared to the model trained by all real datasets. 4) has provable guarantees  that the generator could learn the distributed distribution in an \emph{all important} fashion thus is unbiased.We release our AsynDGAN source code at: https://github.com/tommy-qichang/AsynDGAN

\end{abstract}


\vspace{-1em}
\section{Introduction}
\label{sec:intro}
\subsection{The privacy policies and challenges in medical intelligence}
The privacy issue, while important in every domain, is enforced vigorously for medical data. Multiple level of regulations such as HIPAA~\cite{annas2003hipaa,centers2003hipaa,mercuri2004hipaa,gostin2009beyond} and the approval process for the Institutional Review Board (IRB)~\cite{bankert2006institutional} protect the patients' sensitive data from malicious copy or even tamper evidence of medical conditions~\cite{mirsky2019ct}. Like a double-edge sword, these regulations objectively cause insufficient collaborations in health records.
For instance, America, European Union and many other countries do not allow patient data leave their country~\cite{kerikmae2017challenges,seddon2013cloud}. As a result, many hospitals and research institutions are wary of cloud platforms and prefer to use their own server. Even if in the same country the medical data collaborate still face a big hurdle.

\subsection{The restriction of the medical data accessibility}
It's widely known that sufficient data volume is necessary for training a successful machine learning algorithm~\cite{domingos2012few} for medical image analysis. 
However, due to the policies and challenges mentioned above, it is hard to acquire enough medical scans for training a machine learning model. In 2016, there were approximately 38 million MRI scans and 79 million CT scans performed in the United States~\cite{papanicolas2018health}. Even so, the available datasets for machine learning research are still very limited: the largest set of medical image data available to public is 32 thousand~\cite{yan2018deeplesion} CT images, only 0.02\% of the annual acquired images in the United States.
In contrast, the ImageNet~\cite{deng2009imagenet} project, which is the large visual dataset designed for use in visual object recognition research, has more than 14 million images that have been annotated in more than 20,000 categories.

\subsection{Learning from synthetic images: a solution}
In this work, we design a framework using centralized generator and distributed discriminators to learn the generative distribution of target dataset. In the health entities learning context, our proposed framework can aggregate datasets from multiple hospitals to obtain a faithful estimation of the overall distribution. The specific task (e.g., segmentation and classification) can be accomplished locally by acquiring data from the generator. Learning from synthetic images has several advantages:

\textbf{Privacy mechanism}:
The central generator has limited information for the raw images in each hospital. When the generator communicates with discriminators in hospitals, only information about the synthetic image is transmitted. Such a mechanism prohibits the central generator's direct access to raw data thus secures privacy.

\textbf{Synthetic data sharing}: The natural of synthetic data allows the generator to share the synthetic images without restriction. Such aggregation and redistribution system can build a public accessible and faithful medical database. The inexhaustible database can benefit researchers, practitioners and boost the development of medical intelligence.
  
\textbf{Adaptivity to architecture updates}: The machine learning architecture evolves rapidly to achieve a better performance by novel loss functions~\cite{sudre2017generalised,hochberg1964depth}, network modules~\cite{hoffman2016fcn, ronneberger2015u,milletari2016v,qu2019improving} or optimizers~\cite{ruder2016overview, zeiler2012adadelta, mason2000boosting,zhang2019taming,zhanglocal}. 
We could reasonably infer that the recently well-trained model may be outdated or underperformed in the future as new architectures invented. Since the private-sensitive data may be not always accessible, even if we trained a model based on these datasets, we couldn't embrace new architectures to achieve higher performance. Instead of training a task-specific model, our proposed method trains a generator that learns from distributed discriminators. Specifically, we learn the distribution of private datasets by a generator to produce synthetic images for future use, without worrying about the lost of the proprietary datasets.

To the best of our knowledge, we are the first to use GAN to address the medical privacy problem. Briefly, our contributions lie in three folds: (1) A distributed asynchronized discriminator GAN (AsynDGAN) is proposed to learn the real images' distribution without sharing patients' raw data from different datasets. (2) AsynDGAN achieves higher performance than models that learn from real images of only one dataset. (3) AsynDGAN achieves almost the same performance as the model that learns from real images of all datasets.

%
%


\section{Related Work}
\label{sec:related-work}
\subsection{Generative Adversarial Networks (GANs)}
The Generative Adversarial Nets \cite{goodfellow2014generative} have achieved great success in various applications, such as natural image synthesis~\cite{radford2015unsupervised,zhang2017stackgan,brock2018large}, image style translation~\cite{isola2017image,zhu2017unpaired}, image super resolution~\cite{ledig2017photo} in computer vision, and medical image segmentation~\cite{yang2017automatic,xue2018segan}, cross-modality image synthesis~\cite{nie2017medical}, image reconstruction~\cite{yang2017dagan} in medical image analysis. 
The GAN estimates generative distribution via an adversarial supervisor. Specifically, the generator $G$ attempts to imitate the data from target distribution to make the `fake' data indistinguishable to the adversarial supervisor $D$. In AsynDGAN framework, we mainly focus on the conditional distribution estimation due to the nature of health entities learning problems. However, the AsynDGAN framework can be easily adopted into general GAN learning tasks. 

\subsection{Learning with data privacy}
\textbf{Federated Learning}: The federated learning (FL) seeks to collaborate local nodes in the network to learn a globally powerful model without storing data in the cloud. Recently, FL attracts more attention as data privacy becomes a concern for users~\cite{hard2018federated,konevcny2016federated,brisimi2018federated,huang2018loadaboost}. Instead of directly exposing users' data, FL only communicates model information (parameters, gradients) with privacy mechanism so protects users' personal information. In~\cite{agarwal2018cpsgd, jayaraman2018distributed,mcmahan2016communication}, the SGD is shared in a privacy protection fashion. However, communicating gradients is dimension dependent.
Considering a ResNet101~\cite{he2016resnet} with $d=40$ million parameters, it requires at least $170$ mb to pass gradients for each client per-iteration.
Even with compression technique similar to~\cite{agarwal2018cpsgd}, the communication cost is still  non-affordable for large-size networks.

\textbf{Split Learning}: 
The split learning (SL)~\cite{vepakomma2018split} separates shallow and deep layers in deep learning models. The central processor only maintains layers that are several blocks away from the local input, and only inter-layer information is transmitted from local to central. In this way, the privacy is guaranteed because the central processor has no direct access to data. It reduces the communication cost from model-dependent level to cut-layer-dependent layer while protecting data privacy. However, such method does not apply to neural networks with skip connections, e.g., ResNets~\cite{he2016resnet}. 

In AsynDGAN framework, the communication cost in each iteration is free of the dimension $d$. Only auxiliary data (label and masks), `fake' data and discriminator loss are passed between the central processor and local nodes in the network. For a $128\times128$ size gray-scale image, communication cost per-iteration for each node is $8$ mb with batch size $128$. Since the central processor has only access to discriminator and auxiliary data, the privacy of client is secured via separating block mechanism.

In addition, adaptivity is an exclusive advantage of AsynDGAN framework. With rapid  evolution of machine learning methods, practitioners need to keep updated with state-of-the-art methods. However, there will be a high transaction cost to train a new model in a classical distributed learning subroutine. With the AsynDGAN system, one can maintain the generative distribution. Therefore, updating machine learning models can be done locally with the freedom of generating training data. The comparison between FL, SL and our AsynDGAN is shown in Table~\ref{tabcomp}.

\begin{table}[t] 
	\centering
	\scalebox{0.68}{
	\begin{tabular}{lccc}
		\toprule
		& Privacy Mechanism & Data transmission                                    & Adaptivity \\ \midrule
		FL & Randon Noise           & Parameters / Gradients                                  & No         \\ 
		SL    & Data Block             & Cut Layer Gradients                                   & No         \\ 
		AsynDGAN            & Data Block       & \makecell{Fake Data, Auxiliary Variable\\ \& Discriminator Loss} & Yes        \\ \bottomrule
	\end{tabular}
	}
\vspace{0.02in}
\caption{Comparison between different learning strategies.}
\label{tabcomp}
\end{table}


\section{Method}
\label{sec:method}
\subsection{Overview}
\begin{figure*}[h]
\begin{center}
\includegraphics[width=0.9\linewidth,height=5.5cm]{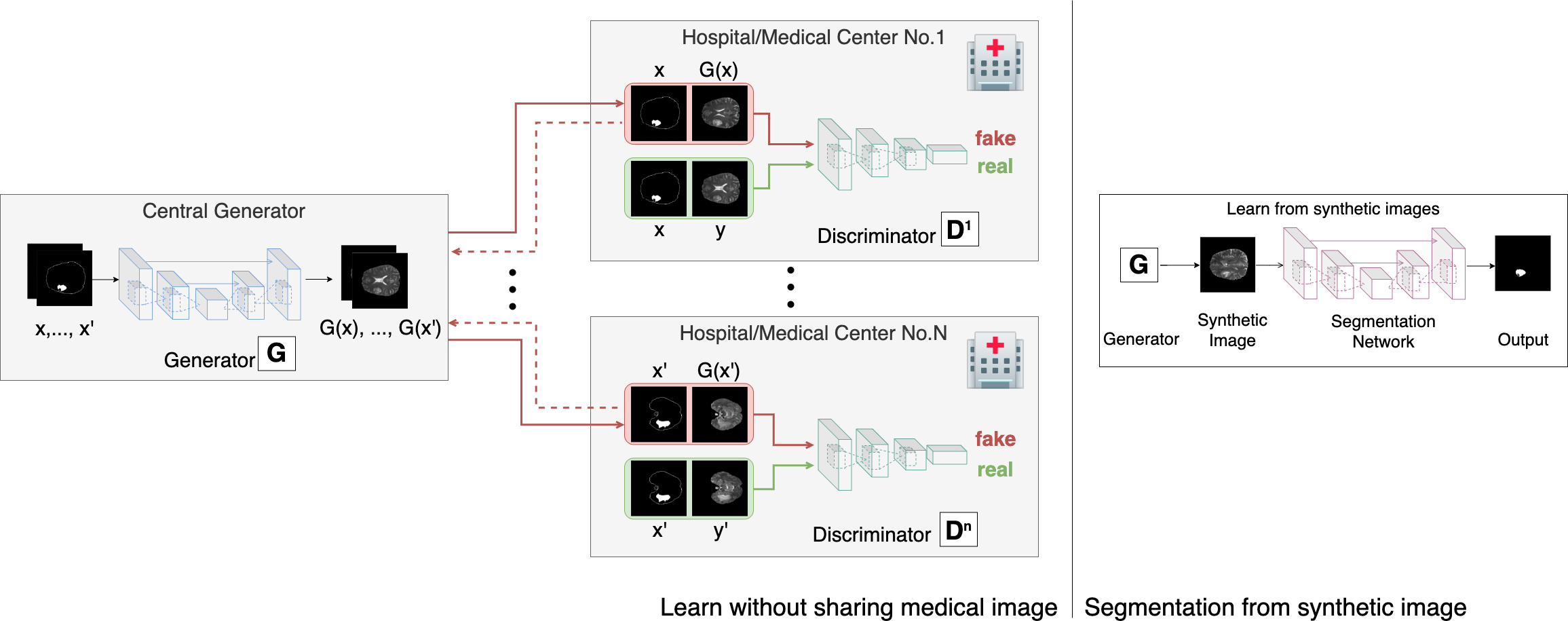}
\end{center}
\caption{The overall structure of AsynDGAN. It contains two parts, a central generator $G$ and multiple distributed discriminators $D^1, D^2, \cdots, D^n$ in each medical entity. $G$ takes a task-specific input (segmentation masks in our experiments) and output synthetic images. Each discriminator learns to differentiate between the real images of current medical entity and synthetic images from $G$. The well-trained $G$ is then used as an image provider to train a task-specific model (segmentation in our experiments).}
\label{arch1}
\end{figure*}

Our proposed AsynDGAN is comprised of one central generator and multiple distributed discriminators located in different medical entities. 
In the following, we present the network architecture, object function and then analysis the procedure of the distributed asynchronized optimization.

\subsection{Network architecture}


An overview of the proposed architecture is shown in Figure~\ref{arch1}.
The central generator, denoted as $G$, takes task-specific inputs (segmentation masks in our experiments) and generates synthetic images to fool the discriminators. The local discriminators, denote as $D^1$ to $D^n$, learn to differentiate between the local real images and the synthetic images from $G$. 
Due to the sensitivity of patients' images, the real images in each medical center may not be accessed from outside. Our architecture is naturally capable of avoiding such limitation because only the specific discriminator in the same medical entity needs to access the real images. In this way, the real images in local medical entities will be kept privately. Only synthetic images, masks, and gradients are needed to be transferred between the central generator and the medical entities. 

The generator will learn the joint distribution from different datasets that belong to different medical entities. Then it can be used as an image provider to train a specific task, because we expect the synthetic images to share the same or similar distribution as the real images. In the experiments, we apply the AsynDGAN framework to segmentation tasks to illustrate its effectiveness. The U-Net~\cite{ronneberger2015u} is used as the segmentation model, and details about $G$ and $Ds$ designed for segmentation tasks are described below.

\subsubsection{Central generator}
For segmentation tasks, the central generator is an encoder-decoder network that consists of two stride-2 convolutions (for downsampling), nine residual blocks~\cite{he2016resnet}, and two transposed convolutions. All non-residual convolutional layers are followed by batch normalization~\cite{ioffe2015batch} and the ReLU activation. All convolutional layers use $3\times3$ kernels except the first and last layers that use $7\times7$ kernels.

\subsubsection{Distributed discriminators}
In the AsynDGAN framework, the discriminators are distributed over $N$ nodes (hospitals, mobile devices). Each discriminator $D_j$ only has access to data stored in the $j$-th node thus discriminators are trained in an asynchronized fashion. For segmentation, each discriminator has the same structure as that in PatchGAN~\cite{isola2016pix2pix}. The discriminator individually quantifies the fake or real value of different small patches in the image. Such architecture assumes patch-wise independence of pixels in a Markov random field fashion \cite{li2016precomputed,isola2017image}, and can capture the difference in geometrical structures such as background and tumors.

\subsection{Objective of AsynDGAN}
The AsynDGAN is based on the conditional GAN~\cite{mirza2014conditiongan}. The objective of a classical conditional GAN is:
\begin{equation}
\begin{aligned}
\min\limits_{G}\max\limits_{D}V(D,G) &= \mathbb{E}_{x\sim s(x)}\mathbb{E}_{y\sim p_{data}(y|x)} [\log D(y|x)]\\
&+\mathbb{E}_{\hat{y}\sim p_{\hat{y}}(\hat{y}|x)} [\log(1-D(\hat{y}|x))]
\end{aligned}
\end{equation}
where $D$ represents the discriminator and $G$ is the generator. $G$ aims to approximate the conditional distribution $p_{data}(y|x)$ so that $D$ can not tell if the data is `fake' or not. The hidden variable $x$ is an auxiliary variable to control the mode of generated data~\cite{mirza2014conditiongan}. In reality, $x$ is usually a class label or a mask that can provide information about the data to be generated. Following previous works~(\cite{mathieu2015deep,isola2016pix2pix}), instead of providing Gaussian noise $z$ as an input to the generator, we provide the noise only in the form of dropout, which applied to several layers of the generator of AsynDGAN at both training and test time.

In the AsynDGAN framework, the generator is supervised by $N$ different discriminators. Each discriminator is associated with a subset of datasets. It is natural to quantify such a setting using a mixture distribution on auxiliary variable $x$. In another word, instead of given a naive $s(x)$, the distributions of $x$ becomes $s(x)=\sum\limits_{j\in[N]} \pi_js_j(x)$. For each sub-distribution, there is a corresponding discriminator $D_j$ which only receives data generated from prior $s_j(x)$. Therefore, the loss function of our AsynDGAN becomes:
\begin{equation}
\begin{aligned}
&\min\limits_{G}\max\limits_{D_1:D_N}V(D_{1:N},G) \\
&= \sum\limits_{j\in [N]} \pi_j \{\mathbb{E}_{x\sim s_j(x)}\mathbb{E}_{y\sim p_{data}(y|x)} [\log D_j(y|x)] \\
&+\mathbb{E}_{\hat{y}\sim p_{\hat{y}}(\hat{y}|x)} [\log(1-D_j(\hat{y}|x))]\}
\end{aligned}
\end{equation}

\subsection{Optimization process}

\begin{figure}[t]
	\vspace{-2em}
	\begin{center}
		\includegraphics[width=5.5cm,height=5.2cm]{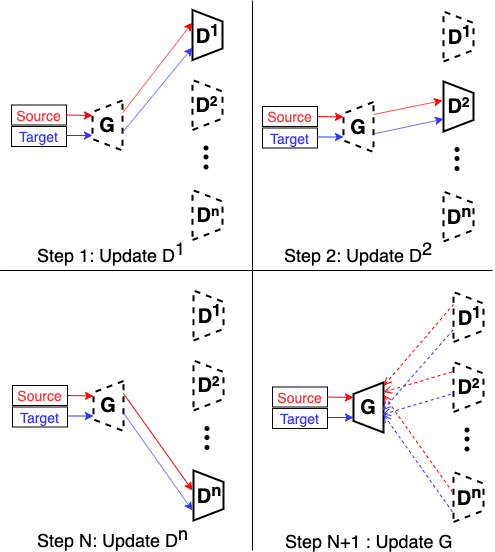}
	\end{center}
	%
	%
	%
	%
	\caption{The optimization process of AsynDGAN. The solid arrows show the forward pass, and the dotted arrows show gradient flow during the backward pass of our iterative update procedure. The solid block indicate that it is being updated while the dotted blocks mean that they are frozen during that update step. Red and blue rectangles are source mask and target real image, respectively.}
	\label{workflow}
\end{figure}

The optimization process of the AsynDGAN is shown in Figure~\ref{workflow}. In each iteration, a randomly sampled tuple $(x, y)$ is provided to the system. Here, $x$ denotes the input label which observed by the generator, and $y$ is the real image only accessible by medical entities. Then the network blocks are updated iteratively in the following order:


\begin{enumerate}[1)]
	\item D-update: Calculating the adversarial loss for $j$-th discriminator $D_j$ and update $D_j$, where $j=1, 2, \cdots, N$.
	\item G-update: After updating all discriminators, $G$ will be updated using the adversarial loss $\sum_{j=1}^N loss(D_j)$.
\end{enumerate}

This process is formulated as Algorithm~\ref{algo1}. We apply the cross entropy loss and in the algorithm and further analyze the AsynDGAN framework in this setting. We stress that the framework is general and can be collaborated with variants of GAN loss including Wasserstein distance and classical regression loss~\cite{arjovsky2017wasserstein,mao2017least}.

\begin{algorithm}[] 
	\caption{\small Training algorithm of AsynDGAN.
	}
	\begin{algorithmic}\label{algo1}
		\FOR{number of total training iterations}
		\FOR{number of interations to train discriminator}
		\FOR{each node $j \in [N]$}
		\STATE{-- Sample  minibatch of of $m$ auxiliary variables $\{x^j_1,...,x^j_m\}$ from $s_j(x)$ and send to generator $G$.}
		\STATE{-- Generate $m$ fake data from generator $G$, $\{\hat{y}^j_1,...,\hat{y}^j_m\}\sim q(\hat{y}|x)$ and send to node $j$.}
		\STATE{-- Update the discriminator by ascending its stochastic gradient:
			\vspace{-0.5em}
			\[\nabla_{\theta_{D_j}} \frac{1}{m} \sum_{i=1}^m \left[
			\log D_j(y_i^j)
			+ \log (1-D_j(G(\hat{y}_i^j)))
			\right].
			\]}
			\vspace{-1.5em}
		\ENDFOR
		\ENDFOR
		\FOR{each node $j \in [N]$}
		\STATE{-- Sample  minibatch of $m$ auxiliary variables $\{x^j_1,...,x^j_m\}$ from $s_j(x)$ and send to generator $G$.}
		\vspace{-0.2em}		
		\STATE{-- Generate corresponding $m$ fake data from generator $G$, $\{\hat{y}^j_1,...,\hat{y}^j_m\}\sim q(\hat{y}|x)$ and send to node $j$.}
		\vspace{-0.2em}
		\STATE{-- Discriminator $D_j$ passes error to generator $G$.}
		\ENDFOR
		\STATE{-- Update $G$ by descending its stochastic gradient:
			\vspace{-0.5em}
			\[	\nabla_{\theta_G} \frac{1}{Nm} \sum_{j=1}^N\sum_{i=1}^m
			\log (1-D_j(G(\hat{y}^j_i))).\]}
			\vspace{-2em}
		\ENDFOR
		\\The gradient-based updates can use any standard gradient-based learning rule. We used momentum in our experiments.
	\end{algorithmic}
	\vspace{-0.3em}
\end{algorithm}

\subsection{Analysis: AsynDGAN learns the correct distribution}

In this section, we present a theoretical analysis of AsynDGAN and discuss the implications of the results. We first begin with a technical lemma describing the optimal strategy of the discriminator.

\begin{lemma}\label{lem1}
	When generator $G$ is fixed,  the optimal discriminator $D_j(y|x)$ is :\\
	\vspace{-0.5em}
	\begin{equation}
	D_j(y|x)=\frac{p(y|x)}{p(y|x)+q(y|x)}
	\end{equation}
\end{lemma}

Suppose in each training step the discriminator achieves its maxima criterion in Lemma \ref{lem1}, the loss function for the generator becomes:\\
\begin{equation*}
\begin{aligned}
\min\limits_{G}V(G)&= \mathbb{E}_{y}\mathbb{E}_{x\sim p_{data}(y|x)} [\log D(y|x)] \\
&+\mathbb{E}_{\hat{y}\sim p_{\hat{y}}(\hat{y}|x)} [\log(1-D(\hat{y}|x))]\\
&=\sum_{j\in[N]} \pi_j\int\limits_{y} s_j(x)\int\limits_{x} p(y|x)\log\frac{p(y|x)}{p(y|x)+q(y|x)}\\
&+q(y|x)\log\frac{q(y|x)}{p(y|x)+q(y|x)} dxdy\\
\end{aligned}
\end{equation*}
Assuming in each step, the discriminator always performs optimally, we show indeed the generative distribution $G$ seeks to minimize the loss by approximating the underlying distribution of data.
\begin{thm}
	Suppose the discriminators $D_{1\sim N}$ always behave optimally (denoted as $D^*_{1 \sim N}$), the loss function of generator is global optimal iff $q(y,x)=p(y,x)$ where the optimal value of $V(G,D^*_{1\sim N})$ is $-\log 4$. 
\end{thm}

\begin{remark}
	While analysis of AsynDGAN loss shares similar spirit with~\cite{goodfellow2014generative}, it has different implications. In the distributed learning setting, data from different nodes are often dissimilar. Consider the case where $\Omega(s_j(x)) \cap \Omega(s_k(y)) =\emptyset, \text{for } k \neq j$, the information for $p(y|x), y\in \Omega(s_j(x))$ will be missing if we lose the $j$-th node. The behavior of trained generative model is unpredictable when receiving auxiliary variables from unobserved  distribution $s_j(x)$.
	The AsynDGAN framework provides a solution for unifying different datasets by collaborating multiple discriminators.
\end{remark}

\section{Experiments}
\label{sec:exp}
In this section, we first perform experiments on a synthetic dataset to illustrate how AsynDGAN learns a mixed Gaussian distribution from different subsets, and then apply AsynDGAN to the brain tumor segmentation task on BraTS2018 dataset~\cite{bakas2018identifying} and nuclei segmentation task on Multi-Organ dataset~\cite{kumar2017dataset}.

\subsection{Datasets and evaluation metrics}
\subsubsection{Datasets}
\paragraph{Synthetic dataset}
The  synthetic dataset is generated by mixing $3$ one-dimensional Gaussian. In another word, we generate $x\in \{1,2,3\}$ with equal probabilities. Given $x$, the random variable $y$ is generated from  $y =y_1{\textbf{1}_{x=1}}+y_2{\textbf{1}_{x=2}}+y_3{\textbf{1}_{x=3}} $ where $\textbf{1}_{event}$ is the indicator function and $y_1\sim \mathcal{N}(-3,2),y_2\sim \mathcal{N}(1,1), y_3\sim\mathcal{N}(3,0.5)$. Suppose the generator learns the conditional distribution of $y$: $p(y|x)$ perfectly, the histogram should behave similarly to the shape of the histogram of mixture gaussian.

\paragraph{BraTS2018}

This dataset comes from the Multimodal Brain Tumor Segmentation Challenge 2018~\cite{bakas2017advancing,bakas2018identifying,menze2014multimodal} and contains multi-parametric magnetic resonance imaging (mpMRI) scans of low-grade glioma (LGG) and high-grade glioma (HGG) patients. There are 210 HGG and 75 LGG cases in the training data, and each case has four types of MRI scans and three types of tumor subregion labels. In our experiments, we perform 2D segmentation on T2 images of the HGG cases to extract the whole tumor regions. The 2D slices with tumor areas smaller than 10 pixels are excluded for both GAN training and segmentation phases. In the GAN synthesis phase, all three labels are utilized to generate fake images. For segmentation, we focus on the whole tumor (regions with any of the three labels).

\paragraph{Multi-Organ}
This dataset is proposed by Kumar et al.~\cite{kumar2017dataset} for nuclei segmentation. There are 30 histopathology images of size $1000\times1000$ from 7 different organs. The train set contains 16 images of breast, liver, kidney and prostate (4 images per organ). The same organ test set contains 8 images of the above four organs (2 images per organ) while the different organ test set has 6 images from bladder, colon and stomach. In our experiments, we focus on the four organs that exist both in the train and test sets, and perform color normalization~\cite{reinhard2001color} for all images. Two training images of each organ is treated as a subset that belongs to a medical entity.

\subsubsection{Evaluation metrics}
We adopt the same metrics in the BraTS2018 Challenge~\cite{bakas2018identifying} to evaluate the segmentation performance of brain tumor: Dice score (Dice), sensitivity (Sens), specificity (Spec), and 95\% quantile of Hausdorff distance (HD95). The Dice score, sensitivity (true positive rate) and specificity (true negative rate) measure the overlap between ground-truth mask $G$ and segmented result $S$. They are defined as
\begin{equation}
Dice(G, S) = \frac{2|G \cap S|}{|G| + |S|}
\end{equation}
\begin{equation}
Sens(G, S)=\frac{|G \cap S|}{|G|}
\end{equation}
\begin{equation}
Spec(G, S)=\frac{|(1-G) \cap (1-S)|}{|1-G|}
\end{equation}
The Hausdorff distance evaluates the distance between boundaries of ground-truth and segmented masks:
\begin{equation}
HD(G, S) = \max\{\sup_{x\in\partial G}\inf_{y\in\partial S}d(x, y), \sup_{y\in\partial S}\inf_{x\in\partial G}d(x, y)\}
\end{equation}
where $\partial$ means the boundary operation, and $d$ is Euclidean distance. Because the Hausdorff distance is sensitive to small outlying subregions, we use the 95\% quantile of the distances instead of the maximum as in~\cite{bakas2018identifying}. To simplify the problem while fairly compare each experiment, we choose 2D rather than 3D segmentation task for the BraTS2018 Challenge and compute these metrics on each 2D slices and take an average on all 2D slices in the test set.

For nuclei segmentation, we utilize the Dice score and the Aggregated Jaccard Index (AJI)~\cite{kumar2017dataset}:
\begin{equation}
AJI = \frac{\sum_{i=1}^{n_\mathcal{G}} |G_i \cap S(G_i)|}{\sum_{i=1}^{n_\mathcal{G}}|G_i \cup S(G_i)| + \sum_{k\in K}|S_k|}
\end{equation}
where $S(G_i)$ is the segmented object that has maximum overlap with $G_i$ with regard to Jaccard index, $K$ is the set containing segmentation objects that have not been assigned to any ground-truth object.

\subsection{Implementation details}
In the synthetic learning phase, we use 9-blocks ResNet~\cite{he2016deep} architecture for the generator, and multiple discriminators which have the same structure as that in PatchGAN~\cite{isola2016pix2pix} with patch size $70\times70$. We resize the input image as $286\times286$ and then randomly crop the image to $256\times256$. In addition to the GAN loss and the L1 loss, we also used perceptual loss as described in \cite{Johnson2016Perceptual}. We use minibatch SGD and apply the Adam solver \cite{kingma2014adam}, with a learning rate of 0.0002, and momentum parameters $\beta_1 = 0.5$, $\beta_2 = 0.999$. The batch size we used in AsynDGAN depends on the number of discriminators. We use batch size 3 and 1 for BraTS2018 dataset and Multi-Organ dataset, respectively.

In the segmentation phase, we randomly crop images of 224$\times$224 with a batch size of 16 as input. The model is trained with Adam optimizer using a learning rate of 0.001 for 50 epochs in brain tumor segmentation and 100 epochs in nuclei segmentation. To improve performance, we use data augmentation in all experiments, including random horizontal flip and rotation in tumor segmentation and additional random scale and affine transformation in nuclei segmentation.

\subsection{Experiment on synthetic dataset}
In this subsection, we show that the proposed synthetic learning framework can learn a mixture of Gaussian distribution from different subsets. We compare the quality of learning distribution in $3$ settings: (1) \textbf{Syn-All.} Training a regular GAN using all samples in the dataset.
(2) \textbf{Syn-Subset-n.} Training a regular GAN using only samples in local subset $n$, where $n\in\{1,2,3\}$. (3) \textbf{AsynDGAN.} Training our AsynDGAN using samples in all subsets in a distributed fashion.


\begin{figure}
\centering
\begin{minipage}{0.325\linewidth}
	\centering\includegraphics[width=\linewidth]{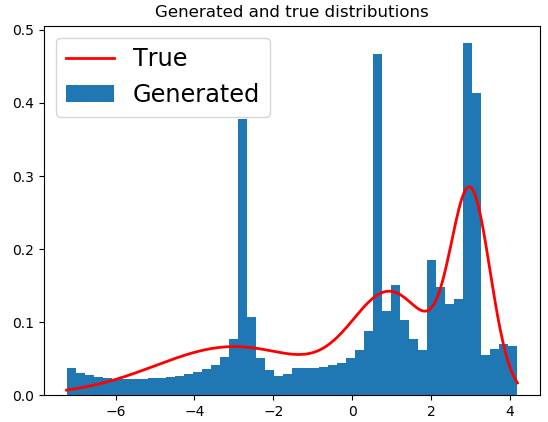} \\ (a) Syn-All
\end{minipage}
\begin{minipage}{0.325\linewidth}
\centering\includegraphics[width=\linewidth]{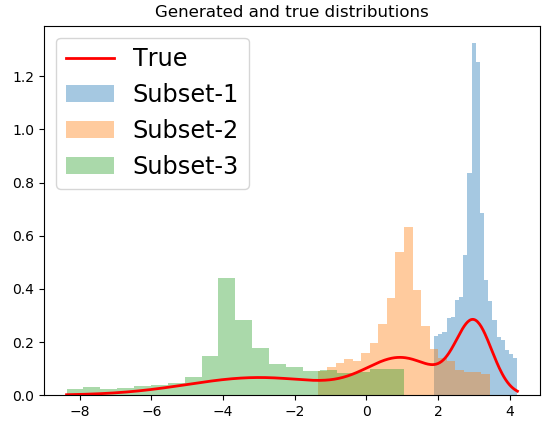} \\ (b) Syn-Subset-n
\end{minipage}
\begin{minipage}{0.325\linewidth}
\centering\includegraphics[width=\linewidth]{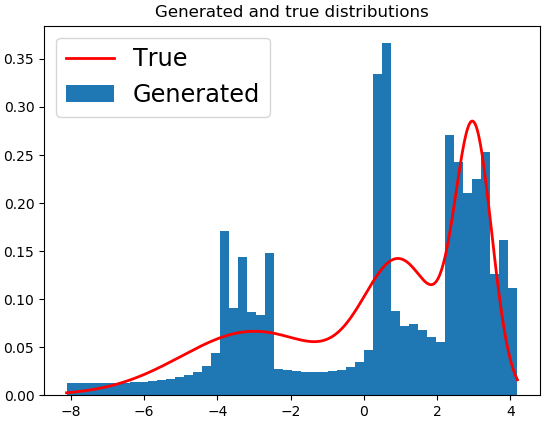} \\ (c) AsynDGAN
\end{minipage}
\caption{Generated distributions of different methods.}
\label{fig:gaussian}
\vspace{-1em}
\end{figure}

The learned distributions are shown in Figure~\ref{fig:gaussian}. In particular, any local learning (indicated in Figure~\ref{fig:gaussian}(b)) can only fit one mode Gaussian due to the restriction of local information while AsynDGAN is able to capture global information thus has a comparable performance with the regular GAN using the union of separated datasets (\textbf{Syn-All}).

\subsection{Brain tumor segmentation}\label{sec:exp:hgg}
In this subsection, we show that our AsynDGAN can work well when there are patients' data of the same disease in different medical entities.
\subsubsection{Settings}
There are 210 HGG cases in the training data. Because we have no access to the test data of the BraTS2018 Challenge, we split the 210 cases into train (170 cases) and test (40 cases) sets. The train set is then sorted according to the tumor size and divided into 10 subsets equally, which are treated as data in 10 distributed medical entities. There are 11,057 images in the train set and 2,616 images in the test set. We conduct the following segmentation experiments:
(1) \textbf{Real-All.} Training using real images from the whole train set (170 cases).
(2) \textbf{Real-Subset-n.} Training using real images from the $n$-th subset (medical entity), where $n=1,2,\cdots,10$. There are 10 different experiments in this category.
(3) \textbf{Syn-All.} Training using synthetic images generated from a regular GAN. The GAN is trained directly using all real images from the 170 cases.
(4) \textbf{AsynDGAN.} Training using synthetic images from our proposed AsynDGAN.
The AsynDGAN is trained using images from the 10 subsets (medical entities) in a distributed fashion.

In all experiments, the test set remains the same for fair comparison. It should be noted that in the \textbf{Syn-All} and \textbf{AsynDGAN} experiments, the number of synthetic images are the same as that of real images in \textbf{Real-All}. The regular GAN has the same generator and discriminator structures as AsynDGAN, as well as the hyper-parameters. The only difference is that AsynDGAN has 10 different discriminators, and each of them is located in a medical entity and only has access to the real images in one subset.

\begin{figure*}[t]
	\vspace{-2em}
	\begin{center}
		\includegraphics[width=0.15\linewidth]{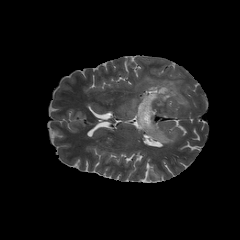}
		\includegraphics[width=0.15\linewidth]{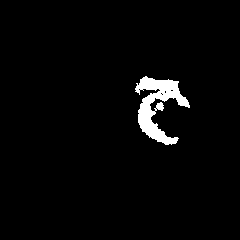}
		\includegraphics[width=0.15\linewidth]{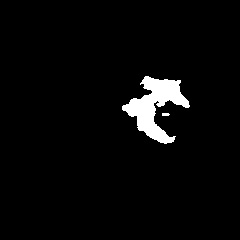}
		\includegraphics[width=0.15\linewidth]{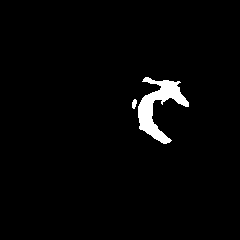}
		\includegraphics[width=0.15\linewidth]{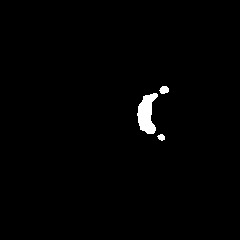}
		\includegraphics[width=0.15\linewidth]{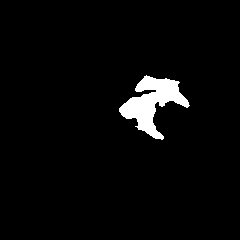} \\ \vspace{0.01in}
		\begin{minipage}{0.15\linewidth}
			\centering\includegraphics[width=\linewidth]{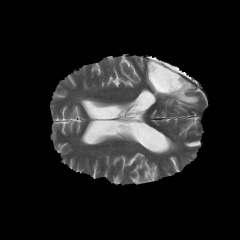} \\ (a) Image
		\end{minipage}
		\begin{minipage}{0.15\linewidth}
			\centering\includegraphics[width=\linewidth]{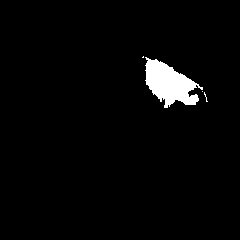} \\ (b) Label
		\end{minipage}
		\begin{minipage}{0.15\linewidth}
			\centering\includegraphics[width=\linewidth]{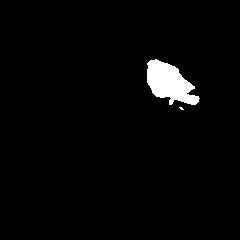}  \\  (c) Real-All
		\end{minipage}
		\begin{minipage}{0.15\linewidth}
			\centering\includegraphics[width=\linewidth]{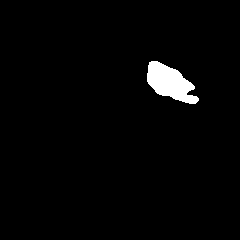} \\ (d) Syn-All
		\end{minipage}
		\begin{minipage}{0.15\linewidth}
			\centering\includegraphics[width=\linewidth]{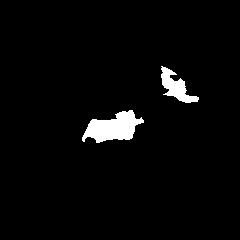} \\ (e) Real-Subset-6
		\end{minipage}
		\begin{minipage}{0.15\linewidth}
			\centering\includegraphics[width=\linewidth]{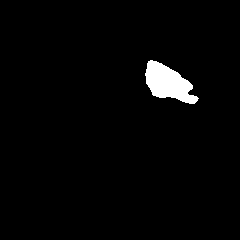} \\ (f) AsynDGAN
		\end{minipage}
	\end{center}
	\caption{Typical brain tumor segmentation results. (a) Test images. (b) Ground-truth labels of tumor region. (c)-(f) are results of models trained on all real images, synthetic images of regular GAN, real images from subset-6, synthetic images of AsynDGAN, respectively.}
	\label{fig:seg:hgg}
\end{figure*}

\begin{figure}[t]
	\begin{center}
		\includegraphics[width=0.3\linewidth]{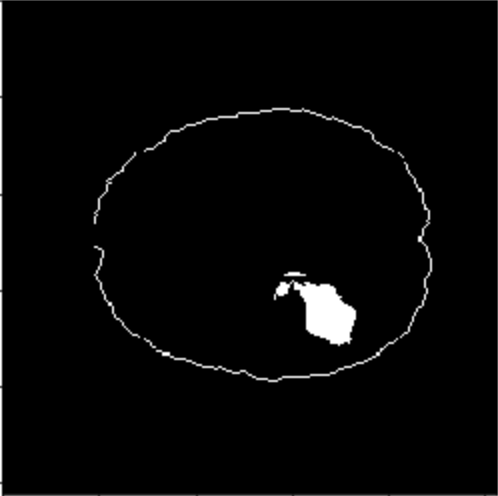}
		\includegraphics[width=0.3\linewidth]{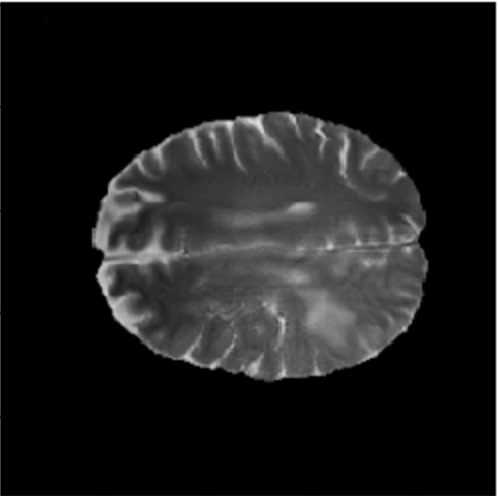}
		\includegraphics[width=0.3\linewidth]{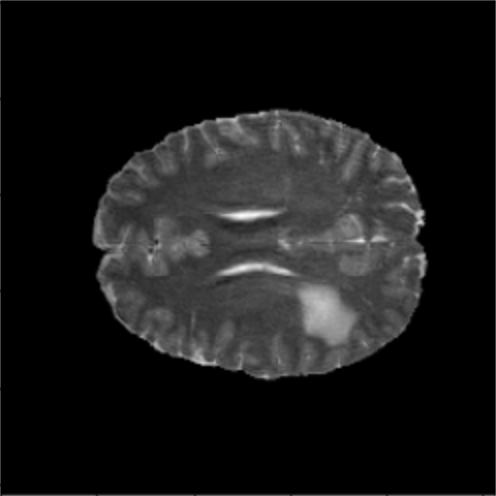}\\ \vspace{0.01in}
		\begin{minipage}{0.3\linewidth}
			\centering\includegraphics[width=\linewidth]{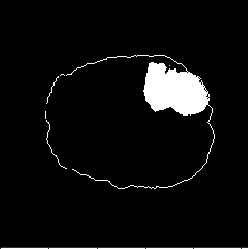} \\ (a) Input
		\end{minipage}
		\begin{minipage}{0.3\linewidth}
			\centering\includegraphics[width=\linewidth]{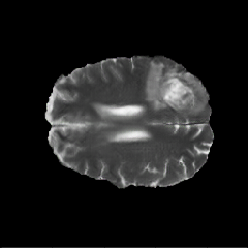} \\ (b) AsynDGAN
		\end{minipage}
		\begin{minipage}{0.3\linewidth}
			\centering\includegraphics[width=\linewidth]{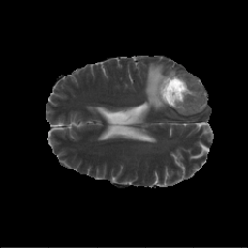}  \\  (c) Real
		\end{minipage}
	\end{center}
	\caption{The examples of synthetic brain tumor images from the AsynDGAN. (a) The input of the AsynDGAN network. (b) Synthetic images of AsynDGAN based on the input. (c) Real images.}
	\label{fig:syn:hgg}
	\vspace{-1em}
\end{figure}

\begin{table}[t]
	\begin{center}
		\begin{tabular}{lcccc}
			\toprule
			Method & Dice $\uparrow$ & Sens $\uparrow$  & Spec $\uparrow$  & HD95 $\downarrow$ \\
			\midrule
			Real-All & 0.7485 & 0.7983	& 0.9955 & 12.85 \\ \midrule
			Real-Subset-1 & 0.5647 &	0.5766 &	0.9945 &	26.90 \\
			Real-Subset-2 & 0.6158 &	0.6333 &	0.9941 &	21.87 \\
			Real-Subset-3 & 0.6660 &	0.7008 &	0.9950 &	21.90 \\
			Real-Subset-4 & 0.6539 &	0.6600 &	0.9962 &	21.07 \\
			Real-Subset-5 & 0.6352 &	0.6437 &	0.9956 &	19.27 \\
			Real-Subset-6 & 0.6844 &	0.7249 &	0.9935 &	21.10 \\
			Real-Subset-7 & 0.6463 &	0.6252 &	0.9972 &	15.60 \\
			Real-Subset-8 & 0.6661 &	0.6876 &	0.9957 &	18.16 \\
			Real-Subset-9 & 0.6844 &	0.7088 &	0.9953 &	18.56 \\
			Real-Subset-10 & 0.6507 &	0.6596 &	0.9957 &	17.33 \\ \midrule
			Syn-All & 0.7114 &	0.7099 &	0.9969 &	16.22 \\ \midrule
			\textbf{AsynDGAN}  & 0.7043 &	0.7295 &	0.9957 &	14.94 \\
			\bottomrule
		\end{tabular}
	\end{center}
	\caption{Brain tumor segmentation results.}	
	\label{tab:hgg}
	\vspace{-1em}
\end{table}

\subsubsection{Results}
The quantitative brain tumor segmentation results are shown in Table~\ref{tab:hgg}. The model trained using all real images (\textbf{Real-All}) is the ideal case that we can access all data. It is our baseline and achieves the best performance. Compared with the ideal baseline, the performance of models trained using data in each medical entity (\textbf{Real-Subset-1$\sim$10}) degrades a lot, because the information in each subset is limited and the number of training images is much smaller.

Our AsynDGAN can learn from the information of all data during training, although the generator doesn't ``see'' the real images. And we can generate as many synthetic images as we want to train the segmentation model. Therefore, the model (\textbf{AsynDGAN}) outperforms all models using single subset. For reference, we also report the results using synthetic images from regular GAN (\textbf{Syn-All}), which is trained directly using all real images. The AsynDGAN has the same performance as the regular GAN, but has no privacy issue because it doesn't collect real image data from medical entities. The examples of synthetic images from AysnDGAN are shown in Figure~\ref{fig:syn:hgg}. Several qualitative segmentation results of each method are shown in Figure~\ref{fig:seg:hgg}.

\begin{figure*}[t]
	\vspace{-2em}
	\begin{center}
		\includegraphics[width=0.15\linewidth]{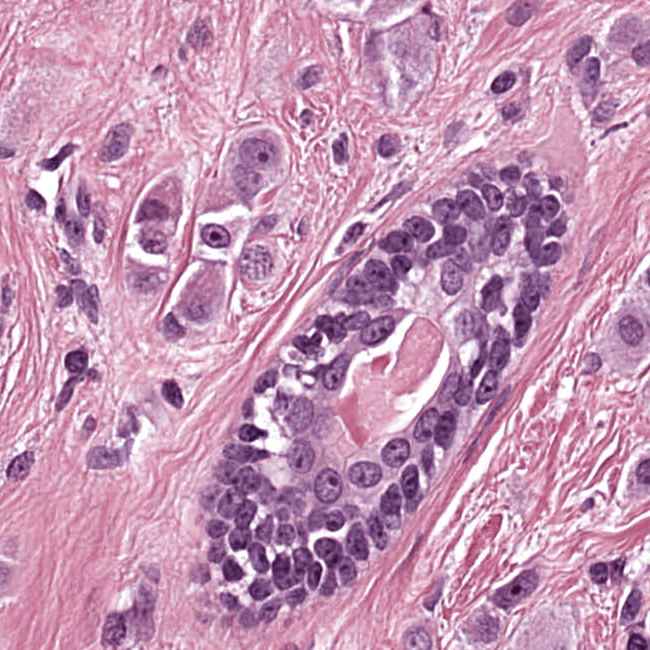}
		\includegraphics[width=0.15\linewidth]{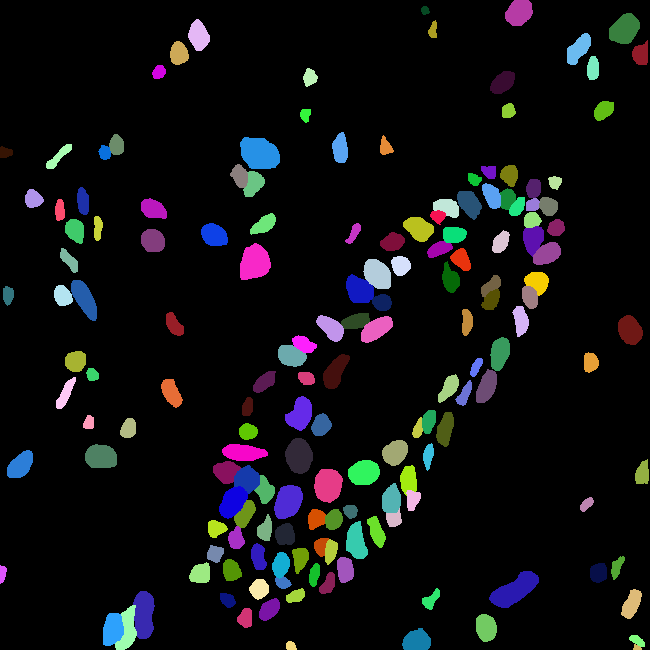}
		\includegraphics[width=0.15\linewidth]{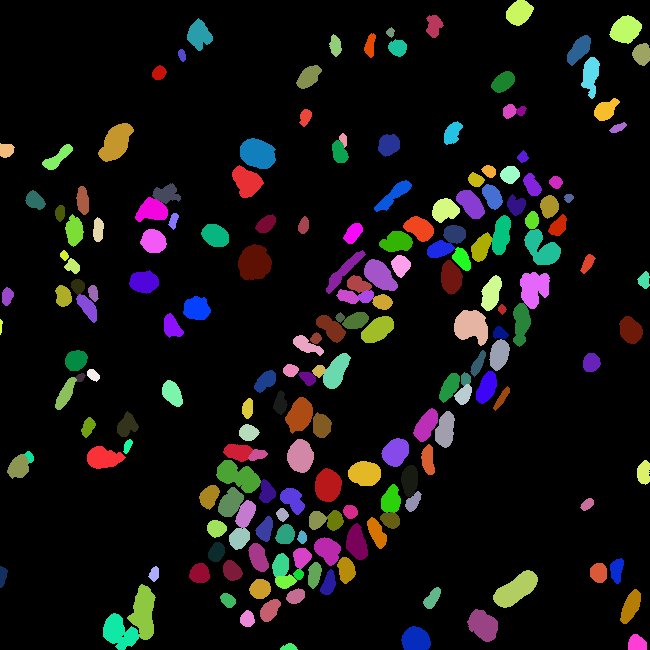}
		\includegraphics[width=0.15\linewidth]{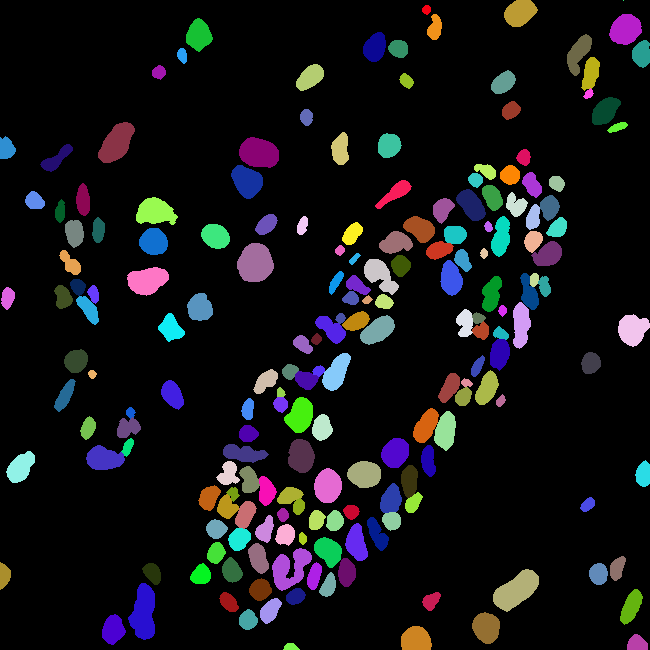}
		\includegraphics[width=0.15\linewidth]{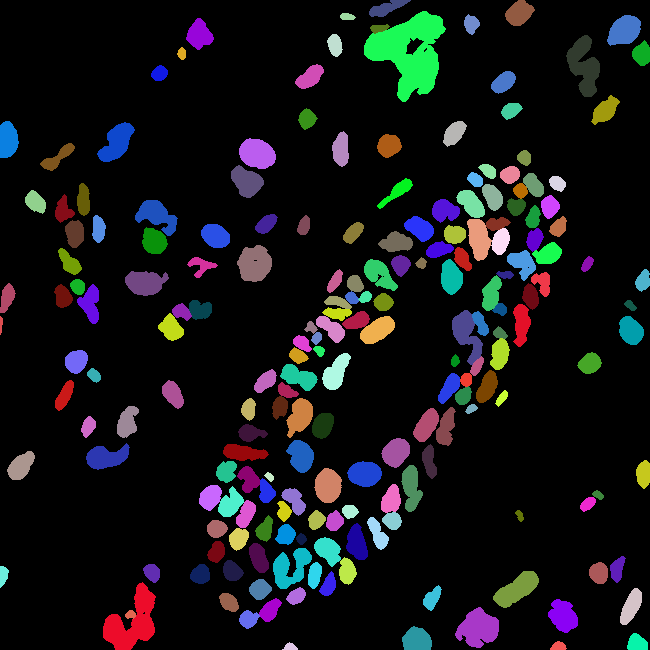}
		\includegraphics[width=0.15\linewidth]{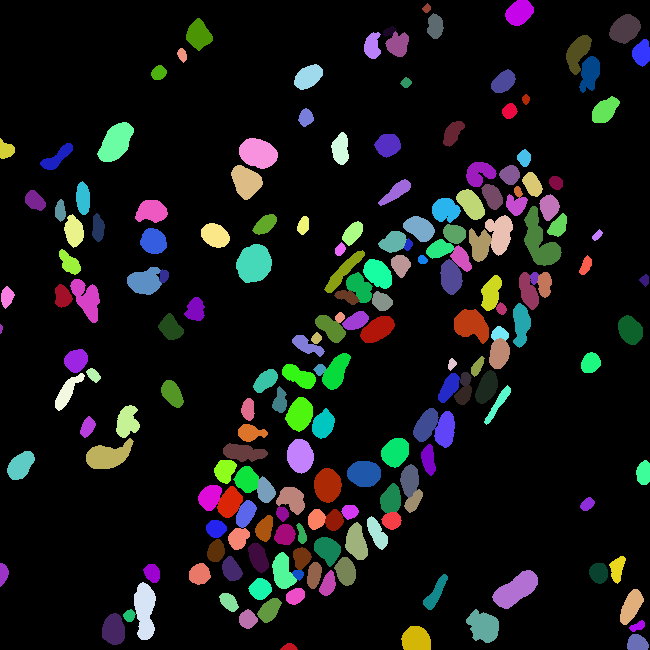} \\ \vspace{0.01in}
		\begin{minipage}{0.15\linewidth}
			\centering\includegraphics[width=\linewidth]{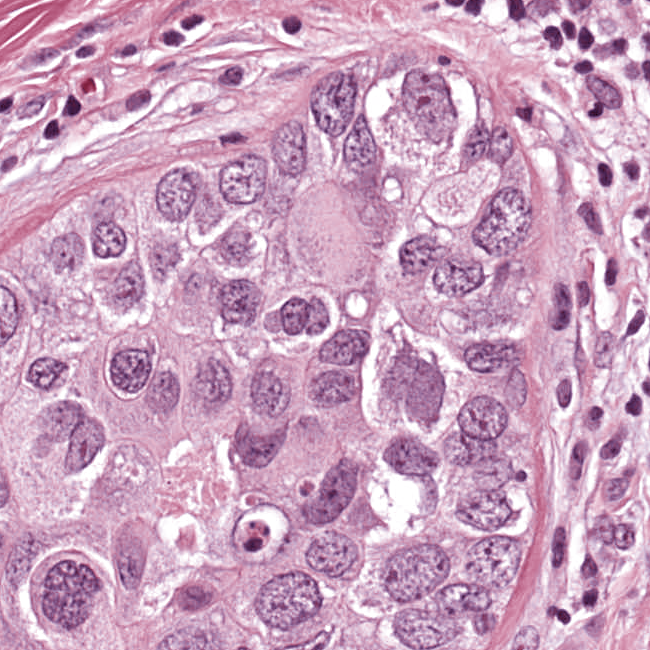} \\ (a) Image
		\end{minipage}
		\begin{minipage}{0.15\linewidth}
			\centering\includegraphics[width=\linewidth]{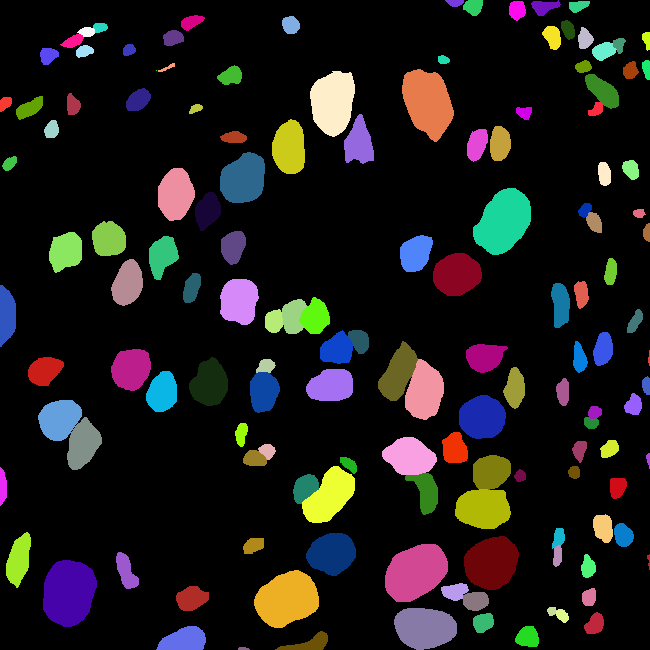} \\ (b) Label
		\end{minipage}
		\begin{minipage}{0.15\linewidth}
			\centering\includegraphics[width=\linewidth]{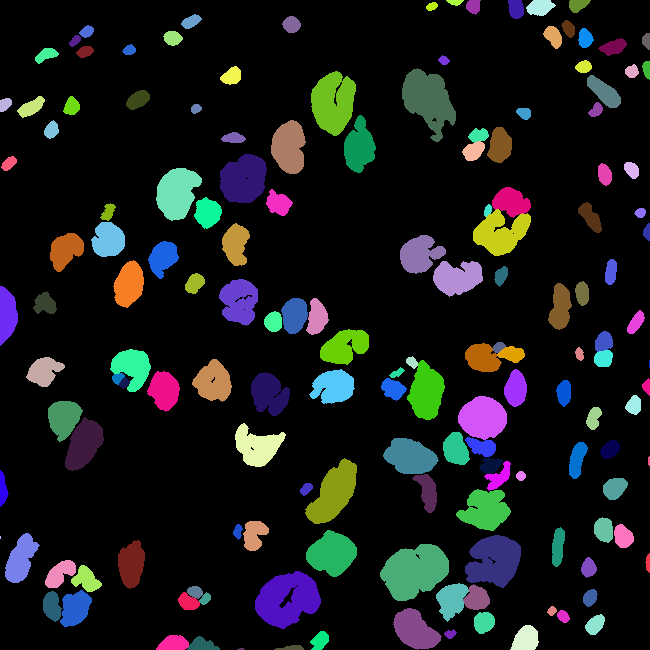}  \\  (c) Real-All
		\end{minipage}
		\begin{minipage}{0.15\linewidth}
			\centering\includegraphics[width=\linewidth]{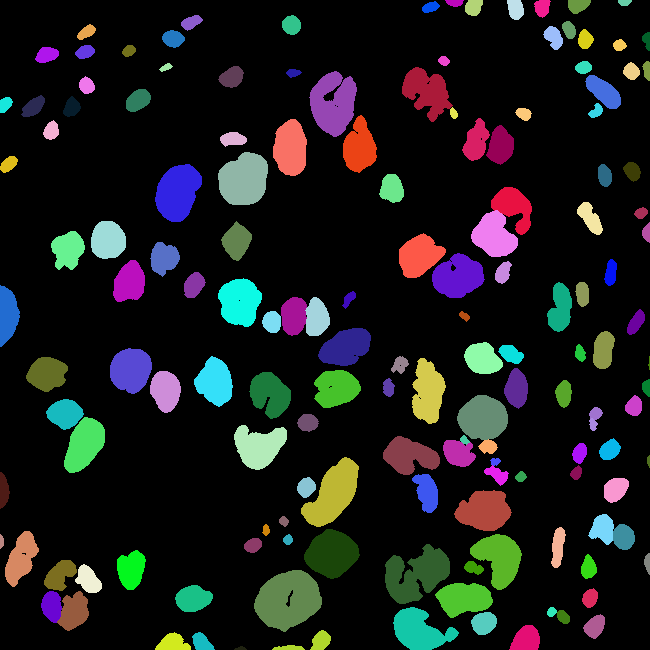} \\ (d) Syn-All
		\end{minipage}
		\begin{minipage}{0.15\linewidth}
			\centering\includegraphics[width=\linewidth]{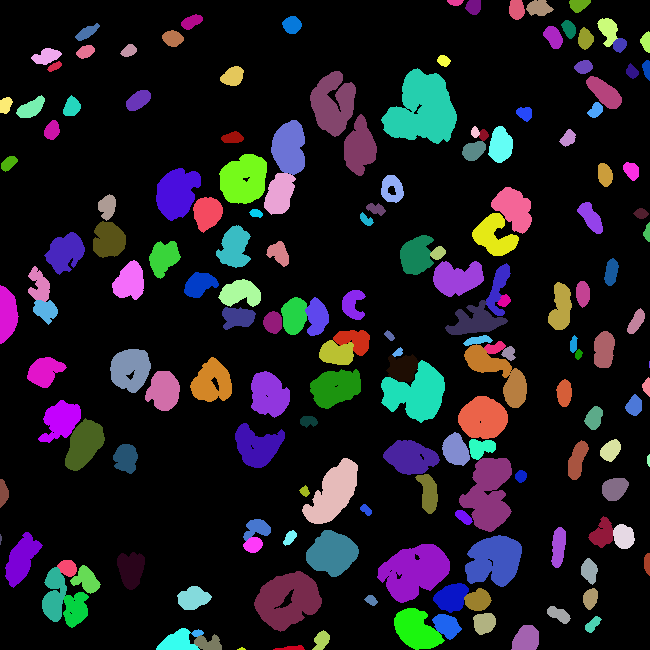} \\ (e) subset-prostate
		\end{minipage}
		\begin{minipage}{0.15\linewidth}
			\centering\includegraphics[width=\linewidth]{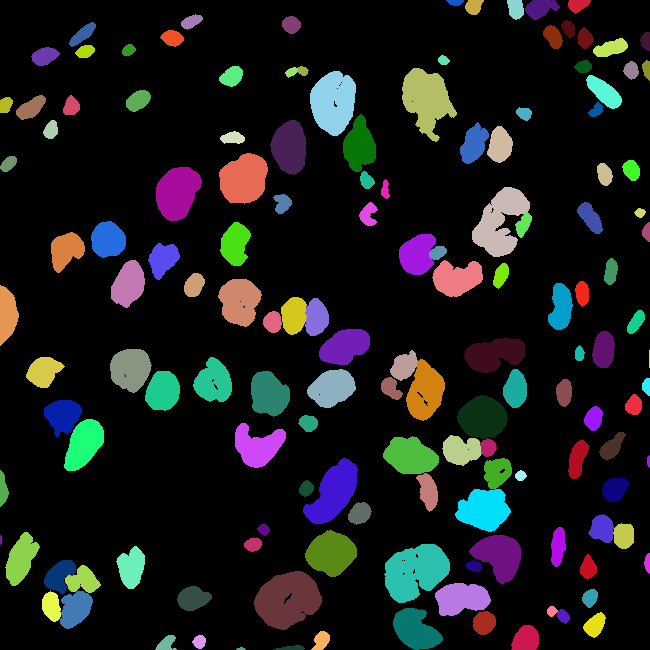} \\ (f) AsynDGAN
		\end{minipage}
	\end{center}
	\caption{Typical nuclei segmentation results. (a) Test images. (b) Ground-truth labels of nuclei. (c)-(f) are results of models trained on all real images, synthetic images of regular GAN, real images from prostate, synthetic images of AsynDGAN, respectively. Distinct colors indicate different nuclei.}
	\label{fig:seg:nuclei}
\end{figure*}

\subsection{Nuclei segmentation}
In this subsection, we apply the AsynDGAN to multiple organ nuclei segmentation and show that our method is effective to learn the nuclear features of different organs. 
\vspace{-1em}
\subsubsection{Settings}
We assume that the training images belong to four different medical entities and each entity has four images of one organ. Similar to Section~\ref{sec:exp:hgg}, we conduct the following experiments:
(1) \textbf{Real-All.} Training using the 16 real images of the train set.
(2) \textbf{Real-Subset-n.} Training using 4 real images from each subset (medical entity), where $n\in\{\text{breast, liver, kidney, prostate}\}$.
(3) \textbf{Syn-All.} Training using synthetic images from regular GAN, which is trained using all 16 real images.
(4) \textbf{AsynDGAN.} Training using synthetic images from the AsynDGAN, which is trained using images from the 4 subsets distributively.
In all above experiments, we use the same organ test set for evaluation.

\begin{table}[t]
	\vspace{-0.7em}
	\begin{center}
		\begin{tabular}{lcc}
			\toprule
			Method & Dice $\uparrow$ & AJI $\uparrow$\\
			\midrule
			Real-All & 0.7833 &	0.5608  \\ \midrule
			Real-Subset-breast & 0.7340  &	0.4942 \\
			Real-Subset-liver & 0.7639 & 0.5191 \\
			Real-Subset-kidney & 0.7416 & 0.4848 \\
			Real-Subset-prostate &0.7704 & 0.5370 \\ \midrule
			Syn-All & 0.7856 &	0.5561 \\ \midrule
			\textbf{AsynDGAN}  & 0.7930 & 0.5608 \\
			\bottomrule
		\end{tabular}
	\end{center}
	\vspace{-0.3em}
	\caption{Nuclei segmentation results.}
	\label{tab:nuclei}
	\vspace{-1em}
\end{table}

\begin{figure}[t]
	\begin{center}
		\includegraphics[width=0.3\linewidth]{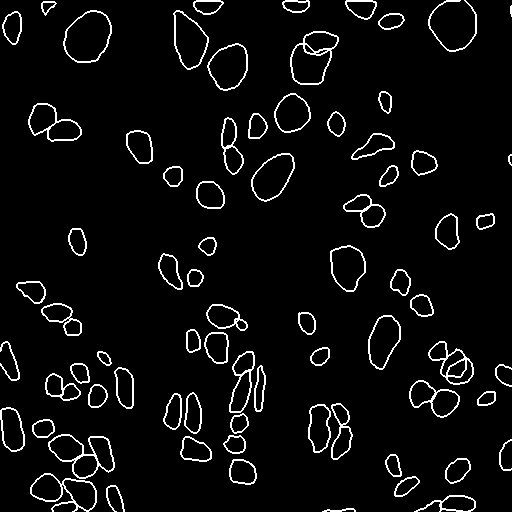}
		\includegraphics[width=0.3\linewidth]{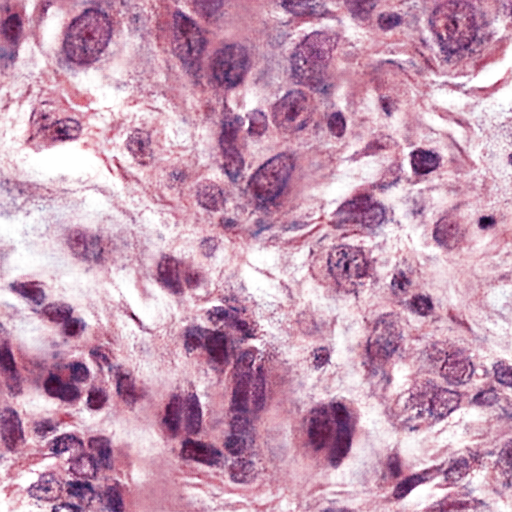}
		\includegraphics[width=0.3\linewidth]{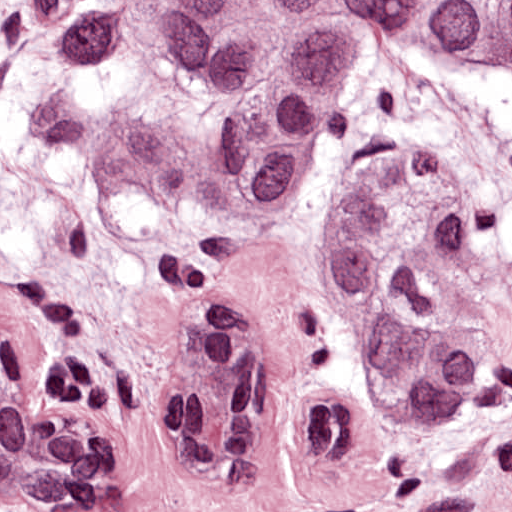}\\ \vspace{0.01in}
		\begin{minipage}{0.3\linewidth}
			\centering\includegraphics[width=\linewidth]{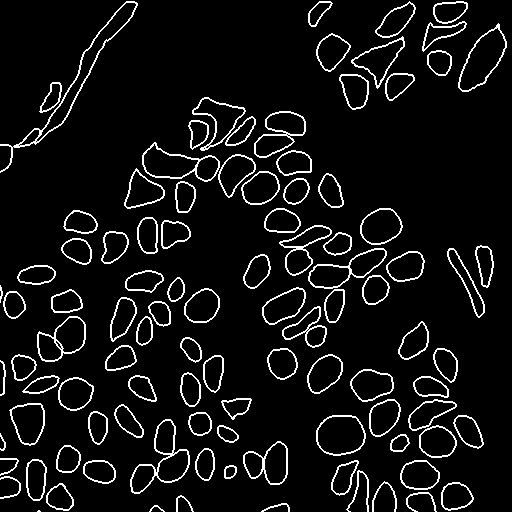} \\ (a) Input
		\end{minipage}
		\begin{minipage}{0.3\linewidth}
			\centering\includegraphics[width=\linewidth]{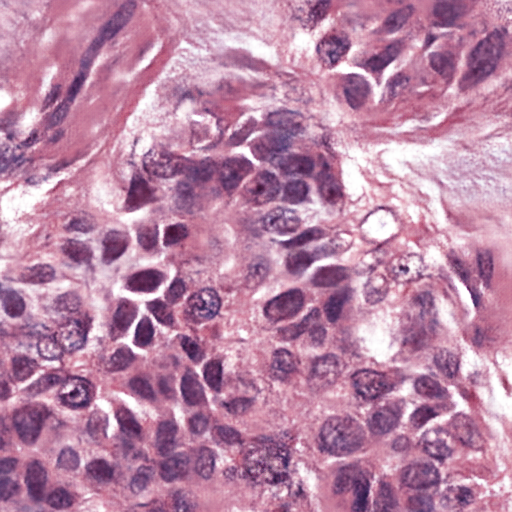} \\ (b) AsynDGAN
		\end{minipage}
		\begin{minipage}{0.3\linewidth}
			\centering\includegraphics[width=\linewidth]{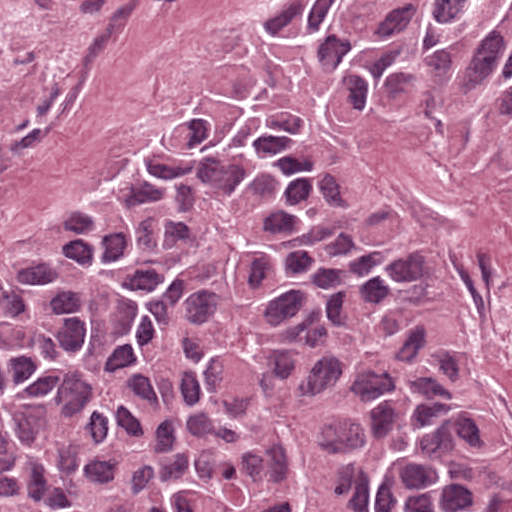}  \\  (c) Real
		\end{minipage}
	\end{center}
	\vspace{-0.5em}
	\caption{The examples of synthetic nuclei images from the AsynDGAN. (a) The input of the AsynDGAN network. (b) Synthetic images of AsynDGAN based on the input. (c) Real images.}
	\label{fig:syn:nuclei}
	\vspace{-0.5em}
\end{figure}
\vspace{-0.5em}
\subsubsection{Results}
The quantitative nuclei segmentation results are presented in Table~\ref{tab:nuclei}. Compared with models using single organ data, our method achieves the best performance. The reason is that local models cannot learn the nuclear features of other organs. Compared with the model using all real images, the AsynDGAN has the same performance, which proves the effectiveness of our method in this type of tasks. The result using regular GAN (\textbf{Syn-All}) is slightly worse than ours, probably because one discriminator is not good enough to capture different distributions of nuclear features in multiple organs. In AsynDGAN, each discriminator is responsible for one type of nuclei, which may be better for the generator to learn the overall distribution. We present several examples of synthetic images from AsynDGAN in Figure~\ref{fig:syn:nuclei}, and typical qualitative segmentation results in Figure~\ref{fig:seg:nuclei}.

\section{Conclusion}
\label{sec:conclude}
In this work, we proposed a distributed GAN learning framework as a solution to the privacy restriction problem in multiple health entities. Our proposed framework applies GAN to aggregate and learns the overall distribution of datasets in different health entities without direct access to patients' data. The well-trained generator can be used as an image provider for training task-specific models, without accessing or storing private patients' data. Our evaluation on different datasets shows that our training framework could learn the real image's distribution from distributed datasets without sharing the patient's raw data. In addition, the task-specific model trained solely by synthetic data has a competitive performance with the model trained by all real data, and outperforms models trained by local data in each medical entity.

\section*{Acknowledgements}
We thank anonymous reviewers for helpful comments. This work was partially supported by ARO-MURI-68985NSMUR, NSF-1909038, NSF-1855759, NSF-1855760, NSF-1733843, NSF-1763523, NSF-1747778 and NSF-1703883.

{\small
\bibliographystyle{ieee_fullname}
\bibliography{main}
}

\clearpage
\onecolumn
	\section{Appendix: AsynDGAN learns the correct distribution}
	In this section we present an analysis of AsynDGAN and discuss the implications of the results. We show that the AsynDGAN is able to aggregates multiple separated data set and learn generative distribution in an \emph{all important} fashion.
	We first begin with a technical lemma describing the optimal strategy of the discriminator.
	\begin{lemma}\label{lem1}
		When generator $G$ is fixed,  the optimal discriminator $D_j(y,x)$ is :\\
		\begin{equation}
		D_j(y,x)=\frac{p(y|x)}{p(y|x)+q(y|x)}
		\end{equation}
	\end{lemma}
	\textbf{Proof}:\\
	\begin{equation*}
	\begin{aligned}
	&\max\limits_{D}V(D)=\max\limits_{D_1\sim D_N}\sum \pi_j\int\limits_{x} s_j(x)\int\limits_{y} p(y|x)log D_j(y,x)+q(y|x)log(1-D_j(y,x)) dydx\\
	&\leq \sum \pi_j\int\limits_{x} s_j(x)\int\limits_{y} \max\limits_{D_j} \{p(y|x)log D_j(y,x)+q(y|x)log(1-D_j(y,x)) \}dydx
	\end{aligned}
	\end{equation*}
	by setting $D_j(y,x)=\frac{p(y|x)}{p(y|x)+q(y|x)}$ we can maximize each component in the integral thus make the inequality hold with equality.\qed
	
	Suppose in each training step the discriminator achieves its maxima criterion in Lemma \ref{lem1}, the loss function for the generator becomes:\\
	\begin{equation*}
	\begin{aligned}
	&\min\limits_{G}V(G)= \mathbb{E}_{y}\mathbb{E}_{x\sim p_{data}(x|y) [logD(y,x)]} +\mathbb{E}_{z\sim p_{\hat{y}}\sim(\hat{y}|x)} [log(1-D(z,y))]\\
	&=\sum_{j\in[N]} \pi_j\int\limits_{x} s_j(x) \underbrace{\int\limits_{y} p(y|x)log\frac{p(y|x)}{p(y|x)+q(y|x)}+q(y|x)log\frac{q(y|x)}{p(y|x)+q(y|x)} dydx}_{\text {To be analyzed in Lemma \ref{lem2}}}
	\end{aligned}
	\end{equation*}
	
	\begin{lemma} \label{lem2}
		Let $a(y)$ and $b(y)$ be two probability distributions s.t. support  $\Omega(a)\subset \Omega(b)$, the loss function $L(a)=\int\limits_{y} a(y)log\frac{a(y)}{a(y)+b(y)}+b(y)log\frac{b(y)}{a(y)+b(y)} dy \geq  -2 log2$. The inequality holds iff $b(y)=a(y)$.
	\end{lemma}
	
	\textbf{Proof}:\\
	Let $\lambda$ be the Lagrangian multiplier. 
	\begin{equation}
	\begin{aligned}
	L(a,\lambda)=\int\limits_{y} a(y)log\frac{a(y)}{a(y)+b(y)}+b(y)log\frac{b(y)}{a(y)+a(y)} + \lambda a(y) \;\;dy -\lambda
	\end{aligned}
	\end{equation}
	By setting $\frac{\partial L}{\partial a} =0$ we have $log\frac{a(y)}{a(y)+b(y)}=\lambda$ holds for all $x$. The fact that $log\frac{a(y)}{a(y)+b(y)}$ is a constant enforces $a(y)=b(y)$. Plugging $a(y)=b(y)$ into loss function we have $L(a)_{|a(y)=b(y)}=-2log(2)$.  \\
	Now we are ready to prove our main result that AsynDGAN learns the correct distribution. Assuming in each step, the discriminator always perform optimally, we show indeed the generative distribution $G$ seeks to minimize the loss by approximating underlying generative distribution of data.
	\begin{thm}
		Suppose the discriminators $D_{1\sim N}$ always behaves optimally (denoted as $D^*_{1 \sim N}$), the loss function of generator is global optimal iff $q(y,x)=p(y,x)$ where the optimal value of $V(G,D^*_{1\sim N})$ is $-log 4$. 
	\end{thm}
	
	\textbf{Proof}:\\
	\begin{equation*}
	\begin{aligned}
	&\min\limits_{\substack{ q(y,x)>0,\\\int\limits_{y} q(y,x)=s(x)}}\sum\limits_{j\in[N]} \pi_j \int\limits_{x}s_j(x)\int\limits_{y} p(y|x)log\frac{p(y|x)}{p(y|x)+q(y|x)}+q(y|x)log\frac{q(y|x)}{p(y|x)+q(y|x)} dydx\\
	&\geq\sum\limits_{j\in[N]} \pi_j \int\limits_{x}s_j(x) \min\limits_{\substack{ q(y|x)>0,\\\int\limits_{y} q(y|x)=1}}\int\limits_{y} p(y|x)log\frac{p(y|x)}{p(y|x)+q(y|x)}+q(y|x)log\frac{q(y|x)}{p(y|x)+q(y|x)} dydx\\
	\end{aligned}\\
	\end{equation*}
	By Lemma \ref{lem2}, the optimal condition for minimizing:\\
	\begin{equation*}
	\min\limits_{\substack{ q(y|x)>0,\\\int\limits_{y} q(y|x)=1}}\int\limits_{y} p(y|x)log\frac{p(y|x)}{p(y|x)+q(y|x)}+q(y|x)log\frac{q(y|x)}{p(y|x)+q(y|x)} dydx
	\end{equation*}
	is by setting $q(y|x)=p(y|x), \forall x$. Such choice of $q(y|x)$ makes the inequality holds as an equality. Meanwhile $q(y|x)=p(y|x)$ implies $q(y,x)=p(y,x)$ given the fact that $p,q$ has the same marginal distribution on $y$. By plugging in $q(y|x)=p(y|x)$ we can derive the optimal value of $V(G,D^*_{1\sim N})$ to be $-log4$.
	
	\begin{remark}
		While analysis of AsynDGAN loss shares similar spirit with \cite{goodfellow2014generative}, it has different implications. In the distributed learning setting, data from different nodes are often dissimilar. Consider the case where $\Omega(s_j(x)) \cap \Omega(s_k(x)) =\emptyset, for k \neq j$, the information for $p(y|x), x\in \Omega(s_j(x))$ will be missing if we lose the $j$-th node. The behavior of trained generative model is unpredictable when receiving auxiliary variables from unobserved  distribution $s_j(x)$.
		The AsynDGAN framework provides a solution for unifying different datasets by collaborating multiple discriminators thus can aggregate separated datasets in an \emph{all important} fashion.
	\end{remark}

\end{document}